\documentclass{ws-ijmpd}

\begin{document}

\title{STATIC CHARGED FLUID IN $(2+1)$-DIMENSIONS ADMITTING CONFORMAL KILLING VECTORS}

\author{FAROOK RAHAMAN}
\address{Department of Mathematics, Jadavpur
University, Kolkata 700 032, West Bengal, India\\
rahaman@iucaa.ernet.in}

\author{SAIBAL RAY}
\address{Department of Physics, Government College of
Engineering \& Ceramic Technology, Kolkata 700 010, West Bengal,
India\\  saibal@iucaa.ernet.in}

\author{INDRANI KARAR}
\address{Department of Mathematics, Saroj Mohan
Institute of Technology, Guptipara, West Bangal, India\\
indrani.karar08@gmail.com}

\author{HAFIZA ISMAT FATIMA}
\address{Department of Mathematics, The
University of Lahore, Lahore, Pakistan\\ ismatfatima4@gmail.com}

\author{SAIKAT BHOWMICK}
\address{Department of Mathematics, Jadavpur
University, Kolkata 700 032, West Bengal, India\\
talkwithsaikat@gmail.com}

\author{GOURAB KUMAR GHOSH}
\address{Department of Mathematics, Jadavpur
University, Kolkata 700 032, West Bengal, India\\
gourabghoshph@gmail.com}

\maketitle

\begin{history}
\received{Day Month Year} \revised{Day Month Year} \comby{Managing
Editor}
\end{history}

\begin{abstract}
New solutions for $(2+1)$-dimensional Einstein-Maxwell space-time
are found for a static spherically symmetric charged fluid
distribution with the additional condition of allowing conformal
killing vectors (CKV). We discuss physical properties of the fluid
parameters.  Moreover, it is shown that the model actually
represents two structures, namely  (i) {\it Gravastar} as an
alternative of black hole and (ii) {\it Electromagnetic Mass}
model depending on the nature of the equation state of the fluid.
Here the gravitational mass originates from electromagnetic field
alone.  The solutions are matched with the exterior region of the
Ba$\tilde{n}$ados-Teitelboim-Zanelli (BTZ) type isotropic static
charged black hole as a consequence of junction conditions. We
have shown that the central charge density is dependent on the
value of $M_0$, the conserved mass of the BTZ black hole. This in
turn depends on the black hole event horizon which again is
related to the Hawking radiation temperature of a BTZ black hole.
Thus one may have a clue that the central charge density is
related with Hawking radiation temperature of the BTZ black hole
on the exterior region of the static charged fluid sphere.
\end{abstract}


\section{Introduction}
The solutions of the charged fluid distribution in
$(2+1)$-dimensional gravity have become a subject of considerable
interest. We search for some new solutions admitting conformal
motion of Killing Vectors (CKV). However, to explore a natural
relationship between space-time geometry and matter distribution
for a star it is an usual practice to take into account the
well-known inheritance symmetry. In this paper, we therefore
investigate the solution of Einstein-Maxwell field equations in
$(2+1)$-dimension by using CKV. Matching conditions of the
spherical charged fluid are imposed on charged BTZ type black hole
in $(2+1)$-dimensional space-time
\cite{BTZ1992,Martinez2000,cataldo2006}.

In this connection we note that study of {\it Gravvastar} i.e.
{\it Grav}itational {\it va}cuum {\it star}, has recently received
tremendous impetus. It was proposed by Mazur and Mottola
\cite{Mazur2001,Mazur2004} as an alternative to black holes by
constructing a new type of solution for the endpoint of a
gravitationally collapsing compact star. In the physical
standpoint they extended the concept of Bose-Einstein condensate
to gravitational systems. The original Mazur-Mottola model
\cite{Mazur2001,Mazur2004} contains an isotropic de Sitter vacuum
in the interior whereas the exterior is a Schwarzschild geometry.
The system was separated by a thin shell of stiff matter such that
the configuration of a gravastar has three different regions: (I)
Interior: $0 \leq r < r_1$ , $p = -\rho$; (II) Shell: $r_1 < r <
r_2$ , $p = +\rho$; and (III) Exterior: $r_2 < r$ , $p = \rho =
0$. It was argued by Mazur and Mottola \cite{Mazur2001,Mazur2004}
that the presence of matter on the thin shell is required to
achieve the required stability of systems under expansion by
exerting an inward force to balance the repulsion from within.
However, Usmani et al. \cite{Usmani2011} proposed a charged
gravastar admitting conformal motion which facilitates stability
of a fluid sphere by avoiding gravitational collapse
\cite{Stettner1973,Whitman1981,Felice1995,Sharma2001,Ivanov2002}.
Some other kind of Gravastar solutions are available in the
literature which demand special mention in connection to our
present investigation
\cite{Sharma2011,Rahaman2012a,Rahaman2012b,Rahaman2012c}.

The scheme of the investigation is as follows. In the next Sec. 2,
we provide Einstein-Maxwell field equations: firstly, in the
compactified implicit forms and secondly, in the expanded explicit
forms for isotropic, static, spherically symmetric charged fluid
in $(2+1)$-dimensional space-time. We explain the interior region
of the sphere under CKV  which represents two different structures
of the model namely (i) Gravastar and (ii) Electromagnetic Mass
model. We rewrite the fluid parameters ($\rho$, $p_{r}$) and
metric coefficients ($e^{\gamma}$, $e^{\lambda}$) in the
appropriate forms to attain physical structures in the Sec. 3
whereas the Sec. 4 gives the exterior region and Sec. 5 matches
the interior to the BTZ-type exterior of the sphere of radius $R$.
The Sec. 6 is used as the platform for providing some concluding
remarks.

\section{The Einstein-Maxwell field equations}
The line element for the interior space-time of a static
spherically symmetric charged distribution of matter in
$(2+1)$-dimensions is taken as \cite{Martinez2000,cataldo2006}

\begin{equation}
ds^2 = -e^{\gamma(r)} dt^2 + e^{\lambda(r)} dr^2 + r^2d\theta^2.
\label{eq1}
\end{equation}

To get Einstein-Maxwell equations, we write the Einstein-Hilbert
action coupled to electromagnetism in the following form

\begin{equation}
I = \int d x^3 \sqrt{-g } \left( \frac{R-2\Lambda}{16 \pi}
-\frac{1}{4} F_a^c F_{bc} + L_{m} \right),\label{eq2}
\end{equation}

where $L_{m}$ is the Lagrangian for matter. The variation with
respect to the fundamental tensor yields  the following self
consistent Einstein-Maxwell equations with cosmological constant
$\Lambda$ for a charged perfect fluid distribution:

\begin{equation}
R_{ab} - \frac{1}{2} R g_{ab} +\Lambda g_{ab} = - 8 \pi (T_{ab}^{PF}
+T_{ab}^{EM}).  \label{eq3}
\end{equation}

The explicit forms of the energy momentum tensor components
for the perfect fluid   and electromagnetic fields are given by:

\begin{equation}
T_{ab}^{PF} = (\rho +p) u_i u_k + p g_{ik},         \label{eq4}
\end{equation}

\begin{equation}
T_{ab}^{EM} = -\frac{1}{4 \pi } \left( F_a^c F_{bc} -\frac{1}{4}
g_{ab} F_{cd}F^{cd}\right), \label{eq5}
\end{equation}

where $\rho$, $p$, $u_i$  are, respectively, matter-energy density,
fluid pressure and three velocity  vector of a fluid element. Here,
electromagnetic field $F_{ab}$ is related to current three vector as

\begin{equation}
F^{ab}_{;b} = - 4 \pi J^a, \label{eq7}
\end{equation}

with
\begin{equation}
J^c = \sigma(r) u^c, \label{eq6}
\end{equation}

where $\sigma(r)$ is the proper charge density of the
distribution.

For this study we assume three velocity  as $u_a = \delta_a^t$ and
concerning the electromagnetic field tensor is given by

\begin{equation}
F_{ab}  =  E(r) (\delta_a^t \delta_b^r-\delta_a^r \delta_b^t),
\label{eq8}
\end{equation}

where $E(r)$ is the electric field.

Hence the Einstein-Maxwell field equations with
cosmological constant ($\Lambda < 0$) under the space-time
 metric (\ref{eq1}) can be explicitly provided as
 (rendering $G = c = 1$)

\begin{eqnarray}
\frac{\lambda' e^{-\lambda}}{2r} &=& 8\pi \rho +E^2 +\Lambda,
\label{eq9} \\ \frac{\gamma' e^{-\lambda}}{2r} &=& 8\pi p -E^2
-\Lambda, \label{eq10}  \\ \frac{e^{-\lambda}}{2}\left(\frac{1}{2}
\gamma'^2+\gamma''-\frac{1}{2}\gamma'\lambda'\right) &=& 8\pi p +E^2
-\Lambda, \label{eq11}\\ \sigma(r)& =&
\frac{e^{-\frac{\lambda}{2}}}{4 \pi r } (rE)',\label{eq12}
\end{eqnarray}

where `$\prime$' denotes differentiation with respect to the
radial parameter $r$. The Eq. (\ref{eq12}) yields the  expression
for electric field as

\begin{equation}
E(r) = \frac{4 \pi }{r} \int_0^r r\sigma(r)e^{\frac{\lambda(r)}{2}}
dr= \frac{q(r)}{r}, \label{eq13}
\end{equation}

where $q(r)$ is total charge of the sphere under consideration. The
generalized Tolman-Oppenheimer-Volkov (TOV) equation for a charged
fluid distribution can be written as

\begin{equation}
\frac{1}{2}\left(\rho + p\right)\gamma' + p' = \frac{1}{8 \pi r^2 }
(r^2E^2)', \label{eq14}
\end{equation}

This is the conservation equation in $(2+1)$-dimensions.

We consider the volume charge density
$\sigma(r)e^{\frac{\lambda(r)}{2}}$ (the term inside the integral
sign in the Eq. (\ref{eq13})) in the polynomial function of $r$. Therefore,
we can write it in the following form

\begin{equation}
\sigma(r)e^{\frac{\lambda(r)}{2}} = \sigma_0 r^n, \label{eq15}
\end{equation}

where $n$ is arbitrary constant as polynomial index and the constant
$\sigma_0$ is referred to the central charge density.

By using Eq. (\ref{eq15}) in Eq. (\ref{eq13}) one can get
the solution for $E(r)$ as follows

\begin{equation}
E(r) = \frac{4 \pi \sigma_0}{n+2} r^{n+1}, \label{eq16}
\end{equation}

and consequently, one can write

\begin{equation}
q(r) = \frac{4 \pi \sigma_0}{n+2} r^{n+2}. \label{eq17}
\end{equation}

Now, the Eq. (\ref{eq9}) implies

\begin{equation}
e^{-\lambda(r)} = M(r),  \label{eq18}
\end{equation}

where

\begin{equation}
M(r) = C -16 \pi \int r\rho(r) dr - 2\int rE^2(r) dr- 2\int r
\Lambda dr,\label{eq19}
\end{equation}

is known as the effective gravitational mass of the spherical distribution
which determines the gravitational field outside the sphere. Here $C$
is an integration constant.

\section{Interior region}
Seeking interior solution we assume the inheritance symmetry of
the spacetime under conformal killing vectors (CKV). Here CKVs are
motions along which the metric tensor of a space-time remains
invariant up to a scale factor.

In a given manifold $M$, we can define a global smooth vector field
$\xi$, known as conformal vector field, such that for the metric
$g_{ab}$ it will take the following form

\begin{equation}
\xi_{a;b} =   \psi g_{ab}+ F_{ab}, \label{eq20}
\end{equation}

where $\psi: M \longrightarrow R $ is the smooth conformal function of
$\xi$ and $F_{ab}$ is the conformal bivector of  $\xi$. This is equivalent to

\begin{equation}
 L_\xi g_{ik} = \psi g_{ik}, \label{eq21}
\end{equation}

where $L$ signifies the Lie derivatives along the CKV $\xi^a$.
Actually CKV provide a deeper insight into the spacetime geometry
and facilitate the generation of exact solutions to the Einstein's
field equations in more comprehensive forms. The study of this
particular symmetry in space-time is physically also very
important because it plays a crucial role of discovering
conservation laws and to devise space-time classification schemes.
Also, it is well known that Einstein's field equations are highly
non-linear partial differential equations and by using CKV, one
can reduce easily the partial differential equations to ordinary
differential equations.

The compactified tensorial Eq. (\ref{eq20}) yields the following
expressions
\begin{eqnarray}
&\xi^1 \gamma^\prime =\psi,\nonumber \\
&\xi^0  = { constant},\nonumber \\
&\xi^1  = \frac{\psi r}{2},\nonumber \\
&\xi^1 \lambda ^\prime + 2( \xi^1 )^\prime =\psi, \nonumber\label{eq22}
\end{eqnarray}

which imply
\begin{eqnarray}
e^\gamma  &=& C_0^2 r^2, \label{23}\\
e^\lambda  &=& \left[\frac {C_1} {\psi}\right]^2,  \label{eq24} \\
\psi &=&\frac{2\xi^{1}}{r},\label{eq25}\\
 \xi^i &=& C_2 \delta_0^i +
\left[\frac{\psi r}{2}\right]\delta_1^i, \label{eq26}
\end{eqnarray}

where the non-zero components of the conformal killing vector
$\xi^a$ are $\xi^0$ and $\xi^1$. Here $C_0$, $C_1$ and $C_2$ all are
integration constants.

Now we will study three different cases with different equations
of state.

\subsection{$p=m\rho$}
By using the equation of state

\begin{equation}
p=m\rho, \label{eq27}
\end{equation}

where $0<m<1$ is an equation of state parameter.

From the TOV Eq. (\ref{eq14}), we obtain the expressions for the
fluid density and pressure as follows:

\begin{equation}
\rho=\frac{4\pi \sigma_{o}^{2}}{(n+2)(3m+1+2mn)}r^{2(n+1)}+\frac{A}{m}r^{-\frac{1+m}{m}},\label{eq28}
\end{equation}

\begin{equation}
p=\frac{4\pi
\sigma_{o}^{2}m}{(n+2)(3m+1+2mn)}r^{2(n+1)}+Ar^{-\frac{1+m}{m}},\label{eq29}
\end{equation}

where $A$ is an integration constant.

At this stage, by using Eqs. (\ref{eq16}) and (\ref{eq28}) in Eq. (\ref{eq19}), we get
the expression for the effective gravitational mass as

\begin{equation}
M(r) = C -Dr^{2(n+2)} +\frac{16\pi A}{m-1}r^{\frac{m-1}{m}} -
\Lambda r^2.\label{eq30}
\end{equation}

where the constant $D$ takes the form

\begin{equation}
D =
\frac{16\pi^{2}\sigma_{o}^{2}(2n+5+3m+2mn)}{(n+2)^{3}(3m+1+2mn)}.\label{eq30a}
\end{equation}

Then from Eq. (\ref{eq18}), after substituting the expression for $M(r)$ from Eq. (\ref{eq30}), we get

\begin{equation}
e^{-\lambda}=C-Dr^{2(n+2)}+\frac{16\pi
A}{m-1}r^{\frac{m-1}{m}}-\Lambda r^{2}.\label{eq31}
\end{equation}

Also, from Eqs. (\ref{eq24}) and (\ref{eq31}), we get the following expression for
the conformal factor $\psi$ as

\begin{equation}
\psi=C_{1}\left(C-Dr^{2(n+2)}-\frac{16\pi
A}{m-1}r^{\frac{m-1}{m}}-\Lambda r^{2}\right)^{\frac{1}{2}}.\label{eq32}
\end{equation}

One can note that the fluid pressure and density fail to be
regular at the origin. However, our demand is that the effective
gravitational mass $M(r)$ should always exist as $r\longrightarrow
0$. This automatically implies that the integration constant $A$
should be zero. Also, for a physically acceptable model, the
energy density should be a decreasing function of radial
coordinate $r$. Therefore Eqs. (\ref{eq28}) and (\ref{eq29})
indicate the following restriction of $n$:

\begin{equation}
-2< n < -1. \label{eq33}
\end{equation}

 To be a realistic star model there should have
some definite value of the density at the surface. However,
effective radial pressure should vanish at the surface $r=R$. This
will be taken care of later on.

Recently, Mazur and Mottola \cite{Mazur2001,Mazur2004} proposed a
new type of solution for the endpoint of a gravitational collapse
in the form of cold, dark and compact object which is a possible
stable alternative to black holes for the endpoint of
gravitational collapse. In the following section, we follow
Mazur-Mottola's concept that  the fluid sphere may contain three
different regions with different equations of state.

\subsection{$p=-\rho$}
Due to Mazur-Mottola model \cite{Mazur2001,Mazur2004}, the
spherical  fluid contains an isotropic de Sitter vacuum in the
interior whereas the exterior is a vacuum black hole solution (in
this case, Ba$\tilde{n}$ados-Teitelboim-Zanelli (BTZ) type
isotropic static charged black hole). In the interior region, let
us consider the equation of state
\begin{equation} p=-\rho.
\end{equation}
Using this, we obtain the expressions for the fluid density and
pressure from the TOV Eq. (\ref{eq14}) as follows:

\begin{equation} p=  -\rho = \frac{2 \pi \sigma_0^2}{(n+2)(n+1)}r^{2(n+1)}-H ,
\end{equation}

where $H$ is an integration constant.

Here, we get the following expression for $M(r)$ as well as
$e^{-\lambda}$ as
\begin{equation}
e^{-\lambda}=M(r) = C+L r^{2(n+2)}+ 8\pi H r^2 -\Lambda r^{2}
\end{equation}
where,
\begin{equation}
L= \frac{16 \pi^2 \sigma_0^2}{(n+1)(n+2)^3}
\end{equation}

We note that $\rho$ is decreasing with increasing $r$ and $M(r)$
as well as volume charge density are regular at the origin. Thus,
one should impose the restriction on $n$ as $ n>0 $.

Keeping in mind the notion of gravastar, we consider apart from
the interior region, that the configuration has a region of thin
shell. One may consider that this thin shell region (Fig. 1)
contains matter having equation of state
\begin{equation} p=\rho.
\end{equation}
Here, the radius of the interior region is assumed to be $R$ and
thickness of the shell is $\epsilon$, no matter how small it is.
This means that outer radius of the fluid configuration is
$R+\epsilon$.

\begin{figure}
\centering
\includegraphics[scale=.3]{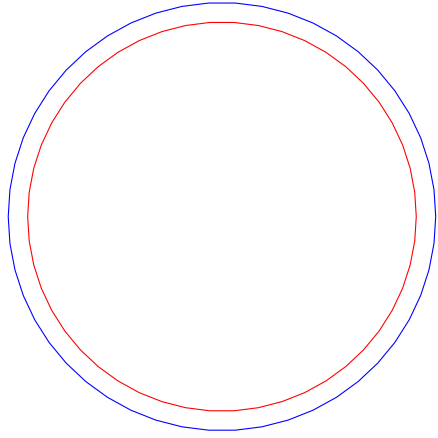}
\caption{The shell with thickness $\epsilon$ contains matter
satisfying the equation of state $p=\rho$ }
    \label{fig:1}
\end{figure}

Using the equation of state (38), we get the following solutions
from the TOV Eq. (\ref{eq14}) and Eq. (19) as
\begin{equation}
p=  \rho = \frac{2 \pi \sigma_0^2}{(n+2)^2}r^{2(n+1)}+ N  r^{-2} ,
\end{equation}

where $N$ is an integration constant.

The metric potential $e^{-\lambda}$ can be written as
\begin{equation}
e^{-\lambda}=M(r) = C- U r^{2(n+2)}- 16 \pi N \ln r -\Lambda
r^{2},
\end{equation}
where
\begin{equation}
U= \frac{32 \pi^2 \sigma_0^2}{(n+2)^3}.
\end{equation}

Imposing the decreasing criteria of density, we take $n<-1$. Here
range of $r$ is $R \leq r \leq R+\epsilon $. Therefore, all the
physical parameters are regular everywhere within the shell.

\section{Exterior region of charged BTZ-type black hole}
Here we have considered that the electro-vacuum exterior solution
($p = \rho = 0$) corresponds to a static, charged BTZ-type black
hole is written in the following form \cite{cataldo2006}

\begin{eqnarray}
ds ^2 = -\left(-M_0 - \Lambda r^2 - Q^2 \ln r \right) dt^2 +\nonumber \\
\left(-M_0 - \Lambda r^2 - Q^2 \ln r\right)^{-1} dr^2 + r^2
d\theta^2,\label{eq34}
\end{eqnarray}

where $Q$ is the total charge and $M_0$ is the conserved mass of the
black hole which is associated with asymptotic invariance under
time-displacements. This mass is given by a flux integral through a
large circle at space-like infinity.

\section{Matching condition}
Let us assume $R$ be the radius of the charged fluid sphere. Then
continuity of the metric functions $g_{tt}$ and $g_{rr}$ across
the surface of charged fluid sphere yields the following results
\cite{Tolman1934}:

\begin{eqnarray}
 C_0^2 R^2 = -M_0-Q^2 \ln R-\Lambda R^2,\label{eq35a}\\
C-DR^{2(n+2)} -\Lambda R^{2} = -M_0-Q^2 \ln R-\Lambda
R^2.\label{eq35b}
\end{eqnarray}

We get the values for integrating constants $C_0$ and $C$ as

\begin{equation}
C_0=\frac{1}{R}(-M_{o}-\Lambda r^{2}-Q^{2}ln R)^{\frac{1}{2}},\label{eq36}
\end{equation}

\begin{equation}
C=DR^{2(n+2)} - M_0 - Q^2 \ln R.\label{eq37}
\end{equation}

 As the effective radial pressure ($ p_{eff} = \frac{\gamma' e^{-\lambda}}{2r}$)
should vanish at the boundary of the charged fluid sphere, therefore from
Eq. (\ref{eq10}), we get

\begin{equation}
\Lambda=-\frac{16(1-m) \pi^2 \sigma_0^2 R^{2n+2}}{(n+2)^2
(3m+1+2mn)}.\label{eq38a}
\end{equation}

Since, $m$ lies between $0$ and $1$, so Eq. (\ref{eq38a}) at once
confirms that the cosmological constant should be negative.
This equation, in turn, yields the radius of the star as

\begin{equation}
R = \left[\frac{-\Lambda (n+2)^2 (3m+1+2mn) }{16(1-m) \pi^2
\sigma_0^2}\right]^{\frac{1}{2n+2}}.\label{eq39}
\end{equation}

From Eq. (30), we get (considering $P_{eff} =
\frac{\gamma' e^{-\lambda}}{2r}=0 $, at the boundary $r=R$)
another expression for $\Lambda$ given by

\begin{equation}
\Lambda = \frac{C}{R^2} -DR^{2(n+1)}.\label{eq38b}
\end{equation}

Definitely, Eqs. (\ref{eq38a}) and (\ref{eq38b}) should be equal.
Then, equating these two equations and after substituting the
value of $D$ from Eq. (\ref{eq30a}), one obtains an expression for
$C$ as follows:

\begin{equation}
C = 16 \pi^2 \sigma_0^2 \left[\frac{( n+3+5m+
3mn)}{(n+2)^3(3m+1+2mn)}\right]R^{2n+4}, \label{eq40}
\end{equation}

which is another form of the equation (\ref{eq37}).

Therefore, after substituting $C$ and $D$ from  Eqs. (\ref{eq40}) and (\ref{eq30a}),
we get the total effective gravitational mass from  Eq. (\ref{eq30}) in the form

\begin{equation}
M(R) = \frac{16 \pi^2 \sigma_0^2 (m-1)}{(n+2)^2(3
m+1+2mn)}R^{2n+4} -\Lambda R^2.\label{eq40a}
\end{equation}

One can note from the above Eq. (\ref{eq40a}) that in the absence
of $\Lambda$ (see Eq. (\ref{eq38a})), the total effective
gravitational mass $M(R)$ becomes negative as constraint on the
equation of state parameter is $0<m<1$. This means inclusion of
$\Lambda$ is unavoidable for the present model to have a usual
positive gravitational mass. Actually, the present model has two
different features depending on the EOS parameter `m'. The first
one, we restrict the EOS parameter `m' as $0<m<1$ which indicates
Electromagnetic origin of the gravitational source, whereas later
on, we relax the restriction on `m' and follow Mazur-Mottola's
concept of gravastar that contains an isotropic de-Sitter vacuum
in the interior. Therefore, we have to take $p=-\rho$ i.e. EOS
parameter having the value $m=-1$.

Again, from the Eqs. (\ref{eq37}) and (\ref{eq40}) we can find out
the expression for the central charge density, $\sigma_0$, in terms of
$Q$, $R$ and $M_0$ as given by

\begin{equation}
\sigma_0^2 = (M_0 + Q^2 \ln R)\left[ \frac{(n+2)^2
(3m+1+2mn)}{16(1-m)\pi^2}\right]R^{-(2n+4)}. \label{eq41}
\end{equation}

Thus it is evident from the above Eq. (\ref{eq41}) that the
central charge density is dependent on the value of $M_0$, the
conserved mass of the BTZ black hole. This in turn depends on the
black hole event horizon which again is related to the Hawking
radiation temperature of a BTZ black hole. Thus one may have a
clue that the central charge density is related with Hawking
radiation temperature of the BTZ black hole on the exterior region
of the static charged fluid sphere.

\section{Physical Analysis}
We provide here first   order derivatives of $\rho$ to show
variational nature of this parameter by assuming from the very
beginning the constraint $A=0$, as explained earlier. This is:

\begin{equation}
 \frac{d\rho}{dr}=\frac{8\pi
\sigma_{o}^{2}(n+1)}{(n+2)(3m+1+2mn)}r^{2n+1}\end{equation}

One can note that $\frac{d\rho}{dr}$ is negative since $n<-1$.
This again supports that $\rho$ is decreasing in nature. In
addition, we would also recall the equation of state, and observe
from the Eq. (\ref{eq26}) that the sound velocity, $v_s^2
=\frac{dp}{d\rho} = m < 1$, which indicates that the model is
realistic.

\section{Conclusion}
In this paper we have presented a new solution set for a
static spherically symmetric fluid distribution under
the frame work of Einstein-Maxwell space-time. As an additional
condition we have allowed CKV to find out inheritance symmetry
of the system of $(2+1)$-dimensional gravity associated with
isotropic fluid. We further consider exterior solution of
BTZ-type static, charged black hole so that our interior solution
can be matched smoothly as a consequence of junction conditions.

Beside the general physical properties these solutions set shows
two special features: (i) {\it gravastar} like configuration, and
(ii) {\it electromagnetic} origin of gravitational sources.

In connection to non-singularity we notice from the Eq.
(\ref{eq30}) that the effective gravitational mass exists at
$r=0$, the constant $C$ (Eq. (\ref{eq40})) being a non-zero
quantity. On the other hand, all the physical parameters $M$, $p$,
$\rho$, $\Lambda$ etc. completely depend on the charge density
$\sigma$, such that for vanishing charge (Eq. (\ref{eq17})) the
charge density vanishes, which in turn makes all the parameters
vanish. This type of solution is known in the literature as {\it
Electromagnetic Mass} model
\cite{Lorentz1904,Feynman1964,Wheeler1962,Wilzchek1999}.
Interestingly our model also represents a charged {\it Gravastar}
as an alternative of black hole as according to Mazur and
Mottola's \cite{Mazur2001,Mazur2004} concept the fluid sphere may
contain three different regions with different equations of state
(Interior: $0 \leq r < r_1 $, ~~~$ p = -\rho $; (II) Shell: $ r_1
< r < r_2 $, ~~~$ p = +\rho $; and (III) Exterior: $ r_2 < r $,
~~~$ p = \rho =0$). The mechanism to achieve this situation due to
inclusion of charge is available in the literature
\cite{Stettner1973,Whitman1981,Felice1995,Sharma2001,Ivanov2002}.

From the solution set of the present studies we observe some of the following
interesting features which raise deeper issues to consider for further investigations:

(1) In Eq. (\ref{eq38a}) we have $\Lambda$, the cosmological
constant to be negative from our calculations. Actually this seems
quite plausible since we have considered the exterior region to be
that of a BTZ-type black hole which has been solved here with a
negative cosmological constant. Here $\Lambda$ has been assumed to
be a constant as was adopted by Einstein for his static
cosmological model. But now there are enough evidence in favour of
an accelerating Universe \cite{Riess1998,Perlmutter1998}, which
demands $\Lambda$ to be a dynamical parameter, a candidate of so
called {\it dark energy}, energy of vacuum. So, a proper follow up
scheme of the present work is to consider further a time varying
phenomenological $\Lambda$ term. Then the radius of the charged
fluid sphere can be shown to vary accordingly. If we can conclude
that the interior solution of the charged fluid sphere is
analogous to that of a charged gravastar admitting inheritance
symmetry then we can have a perfect correspondence of the radius
of the gravastar with dark energy.

(2) We note that our negative cosmological constant $\Lambda$ (Eq.
(\ref{eq38a})) can be expressed in terms of the radius of
curvature, $l = (-\Lambda)^{\frac{1}{2}}$
\cite{BTZ1992,Rubio2008}. The Bekenstein-Hawking entropy
associated with BTZ type of black hole is actually twice the
perimeter of the outer horizon, $S= 4 \pi r_{+}$
\cite{BTZ1992,Rubio2008}. This can be linked with the
characteristic Hawking radiation temperature. So, we see a
possibility that entropy and hence radiation temperature of the
BTZ type charged black hole can be studied which will be done
elsewhere in a future project.

(3) Apart from these we would also like to mention that in our
model we have shown perfect correspondence of the radius of the
gravastar with dark energy. It is well known that the cosmological
constant is now the leading idea to represent dark energy but it
is only applicable if the dark energy EOS parameter relating
pressure and density is $-1$. So that $m=-1$ can be accounted for
this special case, considering dark energy fluid. Thus the range
of $m$ has been provided accordingly. However, other values of $m$
can be accounted for if we can think of other quintessence models
where this EOS parameter of dark energy can vary in accordance.
Since here we have considered $\Lambda$ to be a constant quantity,
so quintessence cannot be taken into account.

However, one drawback in {\it\bf Case 3.1} of our model is that
for $-2<n<-1$, the pressure and density are found to diverge at
$r=0$, and this creates an innocuous possibility. This indicates
that equilibrium configuration requires infinite pressure.
Therefore, it is reasonable to expect that the system would
collapse and cannot be in equilibrium. For a realistic situation,
it is worthwhile to find a non-singular solution of Einstein's
field equations. Work is in progress in this direction.

\section*{Acknowledgments}
FR and SR wish to thank the authorities of the Inter-University
Centre for Astronomy and Astrophysics, Pune, India for providing
the Visiting Associateship under which a part of this work was
carried out. FR is also thankful to UGC, Government of India for
providing financial support under Research Award Scheme. Thanks
are due to anonymous referee for the pertinent issues which helped
us to improve the quality of the manuscript substantially.

\end{document}